\documentclass[aps,pra,floatfix,twocolumn,superscriptaddress,longbibliography]{revtex4-1}
\usepackage[dvips]{graphicx}
\usepackage[english]{babel}
\usepackage{amsmath}
\usepackage{amssymb}
\usepackage{tensor}

\newcommand{\vk}{\mathbf{k}}

\newcommand{\vbr}{\mathbf{r}}
\newcommand{\be}{\begin{eqnarray}}
\newcommand{\ee}{\end{eqnarray}}
\newcommand{\p}{\partial}
\newcommand{\hx}{\hat{x}}
\newcommand{\hy}{\hat{y}}
\newcommand{\dc}{c^{\dagger}}

\def\ket#1{|#1\rangle}
\def\bra#1{\langle #1 |}
\def\ep#1{\langle #1 \rangle}

\begin{document}

\title{The Uhlmann phase of Higher-Order Topological Insulators at Finite Temperature}

\author{Shiyu Chen}
\affiliation{College of Physics, Sichuan University, Chengdu, Sichuan 610064, China}
\email{heyan$_$ctp@scu.edu.cn}

\author{Yan He}
\affiliation{College of Physics, Sichuan University, Chengdu, Sichuan 610064, China}
\email{heyan$_$ctp@scu.edu.cn}

\begin{abstract}
We have studied the finite-temperature topology of higher-order topological insulators (HOTIs) based on the Uhlmann phase, which is a phase angle of the Uhlmann overlap.
%Taking the Benalcazar-Bernevig-Hughes (BBH) model as a typical example of a HOTI, we find that the Uhlmann phase in this case is quantized to $0$ and $\pi$, due to the mathematical properties of the Gamma matrices or the Clifford algebra.
As an example of HOTIs, the Hamiltonian of the Benalcazar-Bernevig-Hughes (BBH) model is constructed from Gamma matrices satisfying the Clifford algebra. This specific algebraic structure underpins the model's higher-order topological properties, including the quantization of the Uhlmann phase to $0$ or $\pi$ .
This quantization enables us to treat the abrupt jumps of the Uhlmann phase as an indication of the nontrivial topological phase of the BBH model at finite temperature. From the disappearance of these jumps, we determine the critical temperature at which the topological transition occurs. For a special choice of parameters, the Uhlmann overlap and the critical temperature can be computed analytically.
\end{abstract}

\maketitle

\section{Introduction}

Recent years have witnessed great progress in the understanding of topological properties of quantum matter, such as topological insulators and topological superconductors~\cite{Kane_TIRev,Zhang_TIRev,Bernevig_book,ShenTI}. The topology of free fermion systems is usually protected by discrete symmetries \cite{Chiu2016}. The nontrivial topology can either be detected by computing the topological index~\cite{TKNN} defined in the bulk system, or by the emergence of edge modes on the system boundary~\cite{hatsugai1993chern}. Despite the rapid developments over the past decade, most works focus on the topology of the ground state at zero temperature. In the real physical world, a quantum system is usually at finite temperature, and the effects of thermal fluctuation are unavoidable. In this case, one has to consider the topology of the density matrix of a mixed state rather than that of a single eigenstate. There have already been many efforts toward understanding the topology of mixed states~\cite{Sjoqvist00,Huang14,Viyuela-1d,Viyuela-review,Diehl,Unanyan20}. This will also be the main focus of the current paper.

The understanding of ground state topology is usually based on the Berry connection~\cite{Berry}, which is a geometric phase acquired by a wavefunction under adiabatic evolution. Starting from the Berry connection, one can determine topological indices such as Chern numbers, Chern characters, and other indicators. In parallel, several pioneering works~\cite{Viyuela-1d,Diehl,Diehl18} have attempted to generalize the concept of the Berry connection from pure states to mixed states. Among these attempts, the Uhlmann connection \cite{Uhlmann,Uhlmann1} is a promising notion, which is defined for the space of amplitudes of full-rank density matrices. The key step in defining the Uhlmann connection is the introduction of a parallel condition in the space of amplitudes, which will be explained in detail in the main text. Based on the Uhlmann connection, several finite-temperature topological indices have been introduced and applied to various one-dimensional and two-dimensional fermion models~\cite{Viyuela-1d,Viyuela-2d,Mera17,HeChern18,YH-BHZ}, spin models~\cite{Galindo21,HouPRA21}, and also quench processes~\cite{HouPRB20,YH-quench}.

In this paper, we will apply the Uhlmann connection to a relatively new type of topological matter called higher-order topological insulators (HOTIs)~\cite{benalcazar2017quantized,okugawa2019second,benalcazar2022chiral,Song-2017,Franca,Yang_2024}. The key feature of HOTIs is that they can support localized corner modes which have codimension 2 or larger. If we put a two-dimensional HOTI system on a square with open boundaries along both directions, there will be four zero modes localized at the corners of the system. The main example of a HOTI that we will consider in this paper is the so-called Benalcazar-Bernevig-Hughes (BBH) model~\cite{benalcazar2017quantized}. This is a two-dimensional (2D) tight-binding model with alternating hopping constants along both directions. The topology of the BBH model is more subtle than that of ordinary Chern insulators. One approach is based on the Wilson loops of the Berry connection, from which the Wannier centers can be obtained. The Wannier center will wind around the entire Brillouin zone for the topological phase of Chern insulators. However, for the BBH model, the Wannier centers will form two separate bands. This motivated the authors of the BBH model to propose a nested Wilson loop based on one of the Wannier bands. Then the electric dipole moments $p_x$ and $p_y$ can be obtained and are shown to be quantized to $0$ and $1/2$ due to the reflection symmetries. The topological phase of the BBH model is indicated by $p_x = p_y = 1/2$. We would like to mention that one can also employ the quadrupole moment as the topological index of the HOTI. The quadrupole moment $q_{xy}$ can also be directly computed in real space~\cite{kang2019many,wheeler2019many}.

We will employ the Uhlmann phase as a topological index at finite temperature for HOTIs, or more specifically the BBH model. Based on the Uhlmann connection, one can parallel transport the amplitude of a mixed state along a closed path in the parameter space. The resulting amplitude is usually different from the amplitude we start with. The Uhlmann overlap is defined as the inner product of these two amplitudes before and after the parallel transport. The phase angle of the Uhlmann overlap is then defined as the Uhlmann phase. We will show that the Uhlmann phase of the BBH model is always quantized to two values, $0$ and $\pi$. Due to this property, we can treat the Uhlmann phase as an indicator of topology at finite temperature. It is found that the topological phase of the BBH model is associated with abrupt jumps of the Uhlmann phase between $0$ and $\pi$. These jumps actually reflect a certain type of modified winding number when one traces around a closed loop in the Brillouin zone. The jumps of the Uhlmann phase disappear at high enough temperature, implying a topological phase transition at a certain critical temperature $T_c$. We can provide a heuristic argument for the causes of these transitions. In a special case, the Uhlmann phase can be analytically computed, which provides a more precise understanding of the transition in that case.

The rest of this paper is organized as follows. In Section~\ref{sec-BBH}, we briefly describe the BBH model and its symmetries. At zero temperature, the topology of the BBH model is described by the method of nested Wannier bands. In Section~\ref{sec-Au}, we introduce the concept of the Uhlmann connection, which is a finite-temperature generalization of the Berry connection. In Section~\ref{sec-U-phase}, we define the Uhlmann phase as the finite-temperature topological index. The numerical results of the Uhlmann phase and the $T_c$ curve for the BBH model are presented and discussed. In this section, we also demonstrate the reason for the quantization of the Uhlmann phase and discuss the causes of the jumps. We also show that $T_c$ can be analytically computed in a certain case. Finally, we briefly conclude in Section~\ref{sec-con}.

\section{The Higher-Order Topological Insulators}
\label{sec-BBH}

In this section, we discuss the prototype example of the higher-order topological insulator, which is the so-called Benalcazar-Bernevig-Hughes (BBH) model.
We will briefly review the symmetry and topological properties of the BBH model at zero temperature. There are several different approaches to characterize the topology of the BBH model. In this paper, we will follow the method of nested Wilson-loops, which is originally proposed in Ref.~\cite{benalcazar2017quantized}.

The Hamiltonian of the BBH model in real space is given as
\be
&&\mathcal{H}=\sum_{\vbr}\Big[m(\dc_{\vbr,1}c_{\vbr,3}+\dc_{\vbr,4}c_{\vbr,2})+m(\dc_{\vbr,1}c_{\vbr,4}-\dc_{\vbr,3}c_{\vbr,2})\nonumber\\
&&\qquad +t(\dc_{\vbr,1}c_{\vbr+\hx,3}+\dc_{\vbr,4}c_{\vbr+\hx,2})\nonumber\\
&&\qquad +t(\dc_{\vbr,1}c_{\vbr+\hy,4}-\dc_{\vbr,3}c_{\vbr+\hy,2})+H.c.\Big]
\label{eq-Ham}
\ee
Here $c_{\vbr,a}^{\dagger}$ and $c_{\vbr,a}$ are the fermion creation and annihilation operators located at the lattice site $\vbr=(x,y)$. We assume that $x,y=1,\cdots,N$, where $N$ is the total number of lattice sites along each axis. The orbital index $a=1,2,3,4$ denotes the four orbitals in each unit cell. We also use $\hx$ and $\hy$ to represent the unit vectors along the $x$ and $y$ directions. The parameters $m$ and $t$ are the intra-unit-cell and inter-unit-cell hopping constants, respectively.

Under the periodic boundary condition, the BBH model can be transformed into a matrix form in the momentum space as follows
\be
&&\mathcal{H}=\sum_{\vk}\psi^{\dag}_\vk H(\vk)\psi_\vk\nonumber\\
&& H(\vk)=(m+t\cos k_x)\Gamma_4-t\sin k_x\Gamma_3\nonumber\\
&&\quad -(m+t\cos k_y)\Gamma_2-t\sin k_y\Gamma_1
\ee
where $\psi_\vk=(c_{\vk,1},\,c_{\vk,2},\,c_{\vk,3},\,c_{\vk,4})^T$. The $\Gamma$ matrices are defined as $\Gamma_{j}=\tau_{2}\sigma_{j}$ for $j=1,2,3$ and $\Gamma_{4}=\tau_{1}\sigma_{0}$. Here $\tau$ and $\sigma$ are two sets of Pauli matrices acting on different orbital degrees of freedom. Note that these $\Gamma$ matrices are the same as those used in the Dirac equation. They satisfy the anti-commutation relations as $\{\Gamma_a,\Gamma_b\}=2\delta_{ab}$. Due to this relation, it is easy to find the two degenerate energy bands of the BBH model as
\be
E=\pm\sqrt{2m^2+2t^2+2mt(\cos k_x+\cos k_y)}
\ee
In the rest of this paper, we set $t=1$ as an energy unit.

The BBH model respects a number of different symmetries, for example, the reflection symmetries along the $x$ and $y$ directions, which are implemented by the matrices $m_x=\tau_1\sigma_3$ and $m_y=\tau_1\sigma_1$, respectively. It is these reflection symmetries that ensure the polarization and quadrupole moment of the BBH model are quantized. These two reflections also satisfy the anti-commuting relation $\{m_x,\,m_y\}=0$, which is important to protect the corner modes.
The BBH model also satisfies the time-reversal and particle-hole symmetries. The combination of these two gives rise to the chiral symmetry. The chiral symmetry operator is given as $\Pi=\tau_{3}\sigma_{0}$, which is anti-commuting with the Hamiltonian as $\{\Pi,\mathcal{H}(\vk)\}=0$.

\begin{figure}[t]
\centering
\includegraphics[width=0.7\columnwidth]{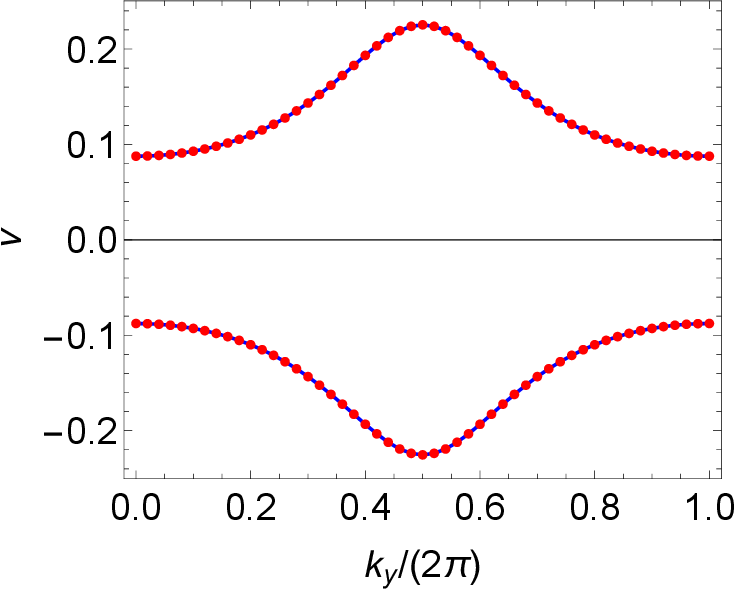}
\caption{The Wannier centers $\nu$ of the BBH model as a function of $k_y$ with $m=0.5$.}
\label{w-band}
\end{figure}

To characterize the topology of the BBH model at zero temperature, we introduce the Wilson loop operator based on the Berry connections. Let us consider the lower band with two degenerate eigenstates $\ket{u_n(\vk)}$ for $n=1,2$. For this occupied band, one can define a non-Abelian Berry connection as follows
\be
(A_\mu)_{m,n}=-i\ep{u_m(\vk)|\frac{\p}{\p k_\mu}|u_n(\vk)}.
\label{eq-Berry}
\ee
with $m,n=1,2$. Roughly speaking, the Berry connection measures the phase difference of the neighboring wave functions in the $\vk$ space. With the Berry connection in hand, the Wilson loop operator is defined as follows
\be
W=\mathcal{P}\exp\Big(i\oint_C A_\mu d k_\mu\Big)
\label{eq-WP}
\ee
Here $C$ denotes a closed loop in $\vk$ space and $\mathcal{P}$ represents the path-ordering, which means that the operator appearing early in this path will apply to the right first. Although $A_\mu$ is not gauge invariant, it is easy to see that the above defined $W$ is gauge invariant and is also a $2\times2$ unitary matrix. For our purpose, we can break the Brillouin zone into a series of loops along $k_x$ axes with fixed $k_y$. Then the resulting Wilson loop $W(k_y)$ is a function of $k_y$. It is well known that the eigenvalues of a unitary matrix are complex number with unit modulus. Therefore, we can write the eigenvalues and eigenstates of $W(k_y)$ as follows
\be
W(k_y)\ket{\nu_j(k_y)}=e^{2\pi i\nu_j(k_y)}\ket{\nu_j(k_y)}\label{W-nu}
\ee
with $j=1,2$ labeling the two different eigenstates. Equation~\ref{W-nu} requires that $\nu_j(k_y)$ is defined only modulo 1. The quantity $\nu_j(k_y)$ is the so-called Wannier center. For the BBH model, the numerical results for the Wannier centers as functions of $k_y$ are plotted in Fig.~\ref{w-band}. The curves of $\nu_j(k_y)$ for $j=1,2$ actually form two separated bands, which are known as the Wannier bands. Now the crucial point is that one can define a topological index for one of the Wannier bands, for example $j=1$. To achieve this, one can first introduce a superposition of the occupied eigenstates according to $\ket{\nu_1(k_y)}$ as
\be
\ket{\tilde{u}(k_y)}=\sum_{n=1}^2\ket{u_n(k_y)}\Big(\ket{\nu_1(k_y)}\Big)_n
\ee
Here $\Big(\ket{\nu_1(k_y)}\Big)_n$ for $n=1,2$ represents the two components of the wave-function $\ket{\nu_1(k_y)}$. Then the nested Wilson loop can be defined as
\be
\widetilde{W}_y=\mathcal{P}\exp\Big(i\oint \tilde{A}_y d k_y\Big),\\
\tilde{A}_y =-i\ep{\tilde{u}(k_y)|\frac{\p}{\p k_y}|\tilde{u}(k_y)}.
\ee
By construction, $\widetilde{W}_y$ is an element of $U(1)$ group. Then the polarization along $y$-axes can be obtained as follows
\be
p_y=\frac{1}{2\pi}\arg \widetilde{W_{y}},\quad 0\leq p_y\leq 1
\ee
Similarly, one can also compute the polarization $p_x$ along $x$-axis. The reflection symmetries require that $p_x=-p_x$ and $p_y=-p_y$, which enforces the quantization of both $p_x$ and $p_y$ to two possible values of $0$ and $1/2$ (modulo 1). The topological phase of the BBH model is indicated by non-zero polarization in both directions $p_x=p_y=1/2$.

\section{Uhlmann connection}
\label{sec-Au}

At finite temperatures, the concept of the Berry connection can be generalized to its mixed-state counterpart, the Uhlmann connection \cite{Uhlmann,Uhlmann1}.
For a mixed state, the physical quantities are all determined by the density matrix $\rho$, which is composed of all the eigenstates of the system. To be more specific, density matrices can usually be expressed as $\rho=\sum_i p_i\ket{u_i}\bra{u_i}$, where $\ket{u_i}$ for $i=1,\cdots,n$ are eigenstates corresponding to the eigenenergies $E_i$, and $p_i$ is the probability of occupying this state. For a mixed state in thermal equilibrium at temperature $T$, $p_i=e^{-E_i/T}$ (the Boltzmann factor). To define the Uhlmann connection, we first introduce the amplitude decomposition of the density matrix as
\be
\rho = w w^{\dagger},\qquad w = \sqrt{\rho}\,U.
\label{eq-ww}
\ee
Here $w$ can be thought of as an analog of the wave function for mixed states. Similar to a wave function, $w$ is not uniquely determined by $\rho$. It is easy to see that an arbitrary choice of the unitary matrix $U$ in Eq.~\eqref{eq-ww} gives rise to the same $\rho$. Therefore, this matrix $U$ represents a $U(n)$ phase factor of the amplitude $w$.

The amplitudes $w$ form a Hilbert space with the Hilbert-Schmidt inner product $(w_1,w_2) = \mbox{Tr}(w_1^\dagger w_2)$, which also defines a metric $|w|^2 = (w,w)$.
Given two different density matrices $\rho_1$ and $\rho_2$, there are infinitely many choices for their amplitudes $w_1 = \sqrt{\rho_1}\,U_1$ and $w_2 = \sqrt{\rho_2}\,U_2$.
In order to fix the relative phase factor between $w_1$ and $w_2$, Uhlmann proposed to minimize the distance $|w_1 - w_2|^2$ over all possible amplitude decompositions \cite{Uhlmann}. He found that the minimization is achieved by requiring that $w_1^\dagger w_2$ is Hermitian and positive definite, which gives rise to the following Uhlmann parallel condition:
\be
w_1^{\dagger}w_2=w_2^{\dagger}w_1>0,
\ee
where the symbol $>0$ means that the above matrix is positive definite.

With this parallel condition, the relative phase factor between $w_1$ and $w_2$ is uniquely determined. To see this, let us first multiply the two sides of the parallel condition to give
\be
w_1^{\dagger}w_2w_2^{\dagger}w_1=U_1^{\dagger}\sqrt{\rho_1}\rho_2\sqrt{\rho_1}U_1
\ee
Taking the square root of the above equation, we find that
\be
w_1^\dag w_2=U_1^{\dagger}\sqrt{\sqrt{\rho_1}\rho_2\sqrt{\rho_1}}\,U_1,
\ee
Then plug in $w_1=\sqrt{\rho_1}U_1$ and $w_2=\sqrt{\rho_2}U_2$, we find that the relative phase factor is given by
\be
U_2U_1^{\dagger}=\sqrt{\rho_2^{-1}}\sqrt{\rho_1^{-1}}\sqrt{\sqrt{\rho_1}\rho_2\sqrt{\rho_1}}.
\label{eq-UU}
\ee
The above result is the Uhlmann connection between $w_1$ and $w_2$ with finite distance. Note that in deriving the above formula, we require the density matrix $\rho_1$ and $\rho_2$ to be full-rank matrices in order to obtain their inverse matrices.

It is more convenient to work with an infinitesimal version of the Uhlmann connection. Considering two nearby density matrices $\rho_1=\rho(\vk)$ and $\rho_2=\rho(\vk+\Delta\vk)$ in Eq.(\ref{eq-UU}), then the infinitesimal Uhlmann connection can be defined as
\be
 A^U_\mu=\p_\mu U U^{\dagger},
\ee
where $\p_\mu=\frac{\p}{\p k_\mu}$. After some straightforward calculations, the explicit result of the Uhlmann connection can be written as
\be
A^U_\mu&=&\sum_{i,j}\ket{u_i}\bra{u_i}\frac{[\p_\mu\sqrt{\rho},\,\sqrt{\rho}]}{p_i+p_j}\ket{u_j}\bra{u_j}\nonumber \\
&=&\sum_{i,j}\frac{(\sqrt{p_i}-\sqrt{p_j})^2}{p_i+p_j}\ket{u_i}\bra{u_i}\p_\mu\ket{u_j}\bra{u_j}.
\label{eq-AU}
\ee
We would like to mention that, if one makes another gauge choice, then $A^U_\mu$ will transform like an ordinary $U(n)$ non-Abelian gauge field. This makes the Uhlmann connection an ideal generalization of the $U(1)$ Berry connection.
Note that the Uhlmann connection $A^U_\mu$ defined in Eq.(\ref{eq-AU}) is \textbf{an} anti-Hermitian matrix, which is different from the Berry connection of Eq.(\ref{eq-Berry}).

\section{The Uhlmann phase of higher order topological insulator}
\label{sec-U-phase}

In this section, we consider the finite temperature topology of HOTIs based on the Uhlmann connections introduced in the previous section. To be specific, we will still take the BBH model as our primary example of a HOTI.
It is well known that the Uhlmann connections behave like a non-Abelian gauge potential, which is very similar to the Berry connection. This fact leads us to construct a gauge-invariant Wilson loop based on the Uhlmann connection as
\be
W^U=\tilde{\mathcal{P}}\exp\Big(\oint_C A^U_{\mu}d k_{\mu}\Big).
\ee
Note that $W^U$ is still a unitary matrix, since $A^U_\mu$ is anti-Hermitian. Here $C$ represents a closed loop in the $\vk$ space and $\tilde{\mathcal{P}}$ represents the anti-path ordering, which is reverse of the path order. We emphasize that the anti-path ordering is important to obtain nontrivial results later.
To be more specific, we will consider a closed loop with a fixed $k_y$ and $k_x$ varying from $0$ to $2\pi$. We can discretize the $k_x$ axis into $N$ points as $k_{x,j}=\frac{2\pi}{N}(j-1)$ with $j=1,\cdots,N$.
Then the Uhlmann Wilson loop can be written as
\be
W^U(k_y)=U_{1,2}U_{2,3}\cdots U_{N,1}
\ee
where $U_{j,j+1}$ is a short-distance Uhlmann Wilson-line for the interval $[k_{x,j},k_{k,j+1}]$. It can be computed as
\be
U_{j,j+1}\approx\exp\Big(A^U_x(k_{x,j},k_y)\Delta k\Big)\nonumber\\
=\sqrt{\rho_2^{-1}}\sqrt{\rho_1^{-1}}\sqrt{\sqrt{\rho_1}\rho_2\sqrt{\rho_1}}.
\ee
where $\rho_1=\rho(k_{x,j},k_y)$ and $\rho_2=\rho(k_{x,j+1},k_y)$.

\begin{figure}[t]
\centering
\includegraphics[width=0.7\columnwidth]{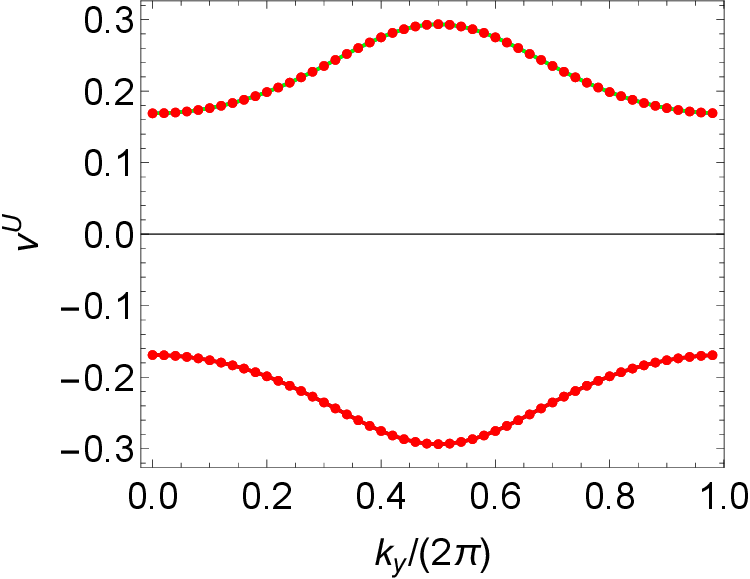}
\caption{The Uhlmann-Wannier center $\nu^U$ of the BBH model as a function of $k_y$ with $m=0.5$.}
\label{U-band}
\end{figure}

By definition, $U_{j,j+1}$ and $W^U$ are 4 by 4 unitary matrices, whose eigenvalues are complex numbers with unit modulus. Therefore, we can define the counterpart of the Wannier center based on the Uhlmann connection similar to Eq.(\ref{W-nu}) as follows
\be
W^U(k_y)\ket{\nu^U_j(k_y)}=e^{2\pi i\nu^U_j(k_y)}\ket{\nu^U_j(k_y)}
\ee
We plot the Uhlmann-Wannier center $\nu^U$ of the BBH model as a function of $k_y$ in Fig.~\ref{U-band}. Although this figure looks similar to Fig.~\ref{w-band}, each curve in Fig.~\ref{U-band} represents two degenerate Uhlmann-Wannier bands.
This degeneracy makes it difficult to construct a nested Wilson loop for each individual band and thus prevents obtaining a nonzero topological index. Due to these difficulties, we need to try another approach to characterize the finite-temperature topology of HOTIs. Inspired by earlier works on one-dimensional topological systems at finite temperatures~\cite{Viyuela-1d}, we introduce the Uhlmann phase, which will be used as a topological index for the BBH model.

To define the Uhlmann phase associated with a closed loop $C$, we first fix a reference point $\vk_0$ on the loop $C$. The amplitude at $\vk_0$ is given by $w_1=\sqrt{\rho(\vk_0)}$. If we parallel transport $w_1$ along the loop $C$ and back to this reference point, the amplitude becomes $w_2=\sqrt{\rho(\vk_0)} W^U$ where $W^U$ is just the Wilson loop based on the Uhlmann connection. Then the Uhlmann phase can be defined as the phase angle of the inner product between $w_1$ and $w_2$, which is given as follows
\be
\Phi^U=\arg\textrm{Tr}(w_1^{\dag}w_2)=\arg\textrm{Tr}(\rho(\vk_0) W^U).
\ee
For later convenience, we also call the inner product $(w_1,w_2)=\textrm{Tr}(\rho(\vk_0) W^U)$ as the Uhlmann overlap.

\begin{figure}[t]
\centering
\includegraphics[width=\columnwidth]{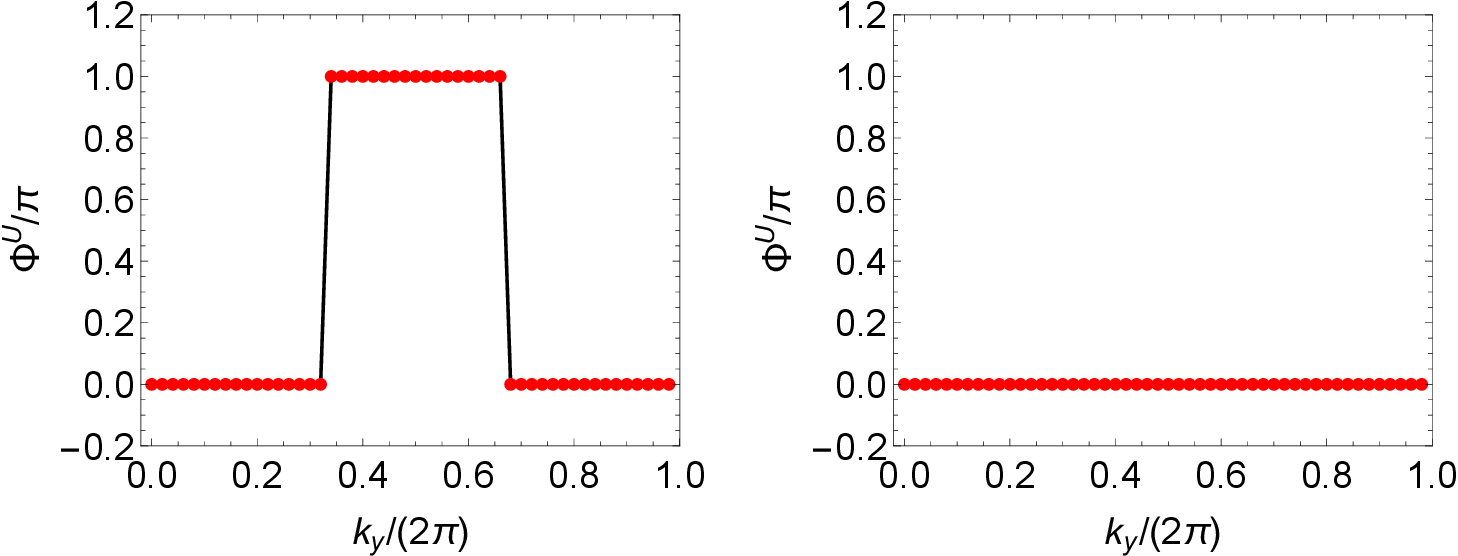}
\caption{The Uhlmann phase $\Phi^U$ of the BBH model as a function of $k_y$ with $m=0.5$. Left panel: $T=0.5$. Right panel: $T=1.5$.}
\label{U-phase}
\end{figure}

\begin{figure}[t]
\centering
\includegraphics[width=0.7\columnwidth]{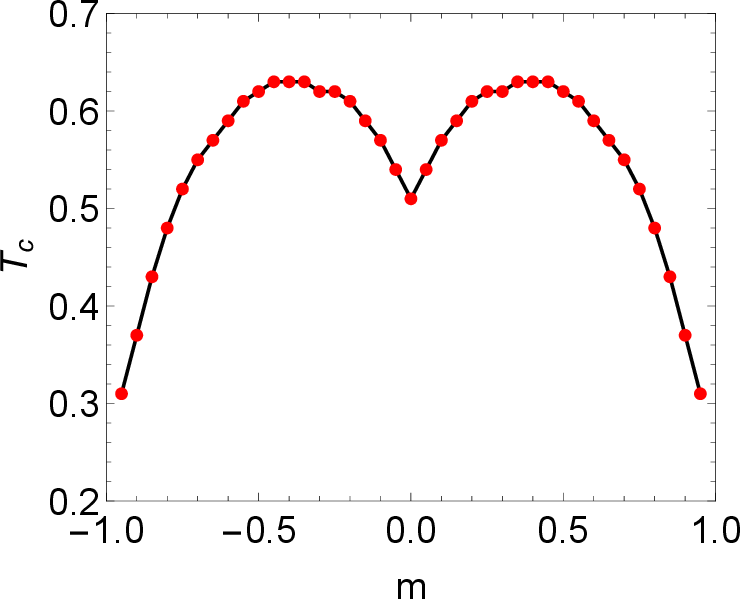}
\caption{The topological transition temperature $T_c$ of the BBH model as a function of $m$.}
\label{fig-Tc}
\end{figure}

Now we can assume that the Brillouin zone is broken into many loops along the $k_x$ axis with fixed $k_y$. For each loop, the Uhlmann phase $\Phi^U$ can be numerically computed. The results of $\Phi^U(k_y)$ for the BBH model as a function of $k_y$ are plotted in Fig.~\ref{U-phase}. We assume $m=0.5$ in the above calculations, which corresponds to the topological phase at $T=0$. In the left panel with $T=0.5$, one can see that $\Phi^U$ shows an abrupt jump from $0$ to $\pi$ at a certain $k_y$ and then jumps back. In the right panel, at a higher temperature $T=1.5$, the Uhlmann phase is always zero. In both cases, $\Phi^U$ is quantized to two possible values: $0$ and $\pi$. Later, we will show that the quantization of $\Phi^U$ is due to a generic feature of a large class of four-band models such as the BBH model. Due to this quantization, we can treat $\Phi^U$ as a topological index of the BBH model at finite temperature. The jumps of $\Phi^U$ in the left panel of Fig.~\ref{U-phase} indicate that the BBH model is in the topological phase at low $T$. For sufficiently high $T$, the jumps disappear, which indicates a trivial phase.

For each given $m$ satisfying $-1 < m < 1$, there exists a critical temperature $T_c$, which separates the low-$T$ topological phase from the high-$T$ trivial phase. The critical temperature can be obtained by raising $T$ until the jumps of $\Phi^U$ disappear. The numerical result of $T_c$ as a function of $m$ is displayed in Fig.~\ref{fig-Tc}. One can see that $T_c$ approaches zero as $m$ approaches $\pm1$, which is consistent with the phase boundary at $T=0$. The $T_c$ curve shows a dip around $m=0$. This is because the average energy gap, roughly given by $\sqrt{2m^2+2t^2}$, reaches a local minimum at $m=0$. The effects of temperature can be thought of as making the energy band blurred. When the temperature is comparable to the energy gap, the two energy bands cannot be sharply separated, and the relative phase angle twisting between these two bands is lost. Therefore, a smaller energy gap implies a lower $T_c$, which is indeed reflected in Fig.~\ref{fig-Tc}.

\subsection{The quantization of the Uhlmann phase}

In this subsection, we will demonstrate why the Uhlmann phase is quantized for the BBH model.
To ease the notation, we rewrite the Hamiltonian of the BBH model as follows
\be
&&H=\sum_{a=1}^4 R_a\Gamma_a \label{eq-BBH}\\
&&R_1=-t\sin k_y,\quad R_2=-(m+t\cos k_y)\nonumber\\
&&R_3=-t\sin k_x,\quad R_4=m+t\cos k_x\nonumber
\ee
Again, the definition of $\Gamma$ matrices is $\Gamma_{a}=\tau_{2}\sigma_{a}$ for $a=1,2,3$ and $\Gamma_4=\tau_1\sigma_0$. We also define $\Gamma_5=\tau_3\sigma_0$.
These five $\Gamma$ matrices form a Clifford algebra by satisfying the anti-commutation relations $\{\Gamma_a,\Gamma_b\}=2\delta_{ab}$ for $a,b=1,\cdots,5$. The BBH model has two degenerate eigen-energy levels $E=\pm R$ with $R=\Big(\sum_{a=1}^4 R_a^2\Big)^{1/2}$. We denote the eigenstates as follows
\be
H\ket{u_{1,2}}=R\ket{u_{1,2}},\quad H\ket{u_{3,4}}=-R\ket{u_{3,4}}
\ee
Clearly, the Boltzmann weights for these two energy levels are
\be
p_{1,2}=\frac{e^{\mp R/T}}{Z},\qquad Z=4\cosh(R/T)
\ee
Denoting the projection operators of the eigen-space corresponding to $\pm R$ as $P_1$ and $P_2$, it is easy to find these two projectors as follows
\be
P_{1,2}=\frac12\Big(I\pm\sum_{a=1}^4\hat{R}_a\Gamma_a\Big)
\ee
here $\hat{R_a}=R_a/R$ and $I$ is a 4 by 4 identity matrix.
Combining the Boltzmann weights and projection operators, one finds that the density matrix is given by
\be
&&\rho=p_1 P_1+p_2P_2=\frac14\Big(I-\sum_{a=1}^4\tanh(\frac{R}{T})\hat{R}_a\Gamma_a\Big)
\ee

According to Eq.(\ref{eq-AU}), the Uhlmann connection $A^U_{\mu}$ can be expressed in terms of $P_{1,2}$ as follows
\be
A^U_{\mu}&=&f_T(P_1\p_\mu P_2+P_2\p_{\mu}P_1)=-\frac12f_T\sum_{a\neq b}\hat{R}_a \p_\mu\hat{R}_b\Gamma_a\Gamma_b\nonumber\\
&=&\frac{i}2f_T\sum_{a<b}\Big(\hat{R}_a \p_\mu\hat{R}_b-\hat{R}_b \p_\mu\hat{R}_a\Big)\Gamma_{ab}
\label{eq-AGa}
\ee
Here we have defined the thermal factor $f_T=1-\dfrac{1}{\cosh(R/T)}$ and $\Gamma_{ab}=i\Gamma_a\Gamma_b$ for $a\neq b$. Since $\Gamma_{ab}=-\Gamma_{ba}$, there are only 6 independent matrices $\Gamma_{ab}$. They also satisfy $\Gamma_{ab}^2=I$.
In deriving the above equation we also have used the fact that $\sum_a\hat{R}_a \p_\mu\hat{R}_a=0$.

It is easy to see that the 6 Hermitian matrices $\Gamma_{ab}$ are actually the generators of the spinor representation of $SO(4)$. Therefore, the commutators between $\Gamma_{ab}$ can still be expressed as a linear combination of $\Gamma_{ab}$ themselves. More explicitly, we can rename $\Gamma_{ab}$ as follows
\be
J_a\equiv\pm\epsilon_{abc}\Gamma_{bc}=\tau_0\sigma_a,\quad
K_a\equiv\Gamma_{a4}=\tau_3\sigma_a
\ee
with $a=1,2,3$. Then the commutators between $\Gamma_{ab}$ or $J_a$ and $K_a$ can be expressed as
\be
&&[J_a,J_b]=2i\epsilon_{abc}J_c,\\
&&[K_a,K_b]=2i\epsilon_{abc}J_c,\\
&&[J_a,K_b]=2i\epsilon_{abc}K_c
\ee
With the above facts in mind, we now turn to the computation of the Uhlmann Wilson loop defined as follows
\be
&&W^U=\tilde{\mathcal{P}}\exp\Big(\oint_C A^U_x(k_x,k_y) d k_x\Big)\nonumber\\
&&\qquad\approx\prod_{j=1}^N\exp\Big( A^U_x(k_{x,j},k_y)\Delta k\Big)
\ee
This product is quite complicated, because the matrices of $A^U_x(k_{x,j},k_y)$ do not commute with each other for different $k_{x,j}$. To proceed, we invoke the Baker-Campbell-Hausdorff (BCH) formula, which states that for any given two operators such as $A$ and $B$, one has the following identity
\be
e^Ae^B=\exp\Big(A+B+\frac12[A,B]+\cdots\Big)
\ee
where $\cdots$ represents all higher order commutators between $A$ and $B$. Due to the BCH formula and the fact that $A^U_x$ is a linear combination of $J_a$ and $K_a$, one immediately sees that the Uhlmann Wilson loop takes the following form
\be
W^U=\exp\Big[i\sum_{a=1}^3(c_a J_a+d_a K_a)\Big]
\ee
where $c_{a}$ and $d_a$ are certain coefficients depending on $k_y$. We cannot compute the coefficients $c_{a}$ and $d_a$ analytically in general. However, one can deduce that these coefficients must be real, since $W^U$ is a unitary matrix by definition. Because the generators of $J_a$ and $K_a$ are all block diagonalized, the Uhlmann Wilson loop $W^U$ is also a block diagonalized matrix, which can be written as
\be
W^U=\left(
    \begin{array}{cc}
      \exp(i\sum_a r_{1a}\sigma_a) & 0 \\
      0 & \exp(i\sum_a r_{2a}\sigma_a)
    \end{array}
  \right)\label{eq-WU}
\ee
where $r_{1a}=c_a+d_a$ and $r_{2a}=c_a-d_a$. These 2 by 2 blocks can be expanded as
\be
\exp(i\sum_a r_{ja}\sigma_a)=\cos r_j+i\sin r_j\sum_a\hat{r}_{ja}\sigma_a\label{eq-exp}
\ee
for $j=1,2$. Here we defined $r_j=\sqrt{\sum_a r_{ja}^2}$ and $\hat{r}_{ja}=r_{ja}/r_j$.
Substituting Eq.~(\ref{eq-exp}) into Eq.~(\ref{eq-WU}), we finally obtain the result of the Uhlmann Wilson loop $W^U$ as
\be
&&W^U=\frac12(\cos r_1+\cos r_2)I+(\cos r_1-\cos r_2)\Gamma_5\nonumber\\
&&\qquad +\frac i2\sum_a\Big[(\sin r_1\hat{r}_{1a}+\sin r_2\hat{r}_{2a})J_a\nonumber\\
&&\qquad +(\sin r_1\hat{r}_{1a}-\sin r_2\hat{r}_{2a})K_a\Big]
\ee
Note that $\Gamma_a$, $J_a$ and $K_a$ are all traceless matrices. We also have $\mbox{Tr}(\Gamma_a\Gamma_5)=0$ and $\mbox{Tr}(\Gamma_aJ_b)=\mbox{Tr}(\Gamma_aK_b)=0$.
Making use of these facts, we find that the Uhlmann phase is given by
\be
\Phi^U=\arg\mbox{Tr}(\rho_0 W^U)=\arg(\cos r_1+\cos r_2)
\ee
Note that the quantity inside the parenthesis must be a real number. This proves that the Uhlmann phase $\Phi^U$ must be quantized to $0$ or $\pi$ for a Hamiltonian which is a linear combination of $\Gamma_a$ matrices just like Eq.(\ref{eq-BBH}). The mathematical form of the BBH Hamiltonian is partly determined by the mirror reflection and chiral symmetries. In this sense, the quantization of the Uhlmann phase is also due to the symmetries of the model.

\begin{figure}[t]
\centering
\includegraphics[width=0.7\columnwidth]{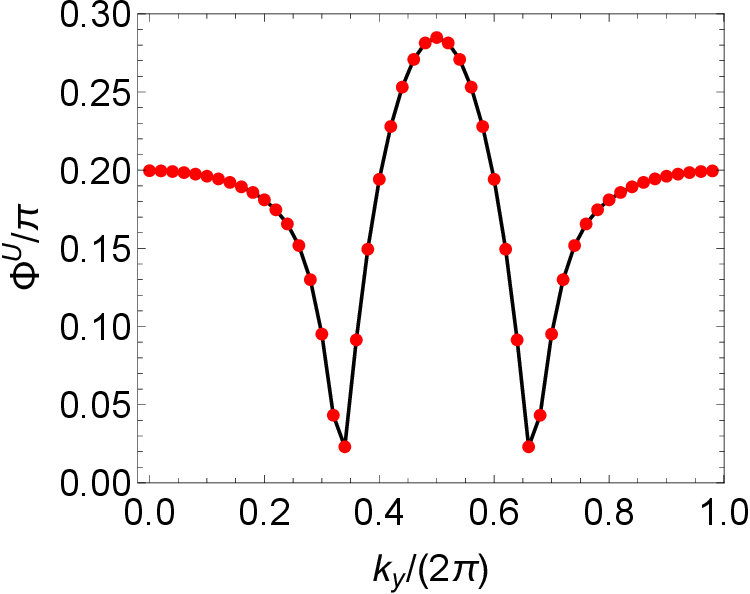}
\caption{The Uhlmann phase $\Phi^U$ of Eq.(\ref{eq-Hp}) as a function of $k_y$ with $m=0.5$ and $T=0.5$.}
\label{U-phase-1}
\end{figure}

One may wonder what will happen to the Uhlmann phase if the Hamiltonian also involves $\Gamma_{ab}$. In this case, our previous arguments are not valid any more, and it is likely that the Uhlmann phase is no longer quantized.  As an example, we consider the following Hamiltonian
\be
H'=H+\sin k_x\,\Gamma_{24}\label{eq-Hp}
\ee
where $H$ is the BBH model Hamiltonian of Eq.(\ref{eq-BBH}). We plot the Uhlmann phase of the above model as a function of $k_y$ in Fig.\ref{U-phase-1}. Indeed, one can see that $\Phi^U$ continuously varies with $k_y$ because of the term of $\Gamma_{24}$.

\subsection{Discussions about the finite $T$ transition}

To understand why $\Phi^U$ jumps below $T_c$, we can make an approximated computation of Uhlmann phases. Let us assume that the 4-component vector $R_a$ with $a=1,\cdots,4$ has two components with much larger magnitude than the other two. To be more specific, let us assume that the dominant components are $R_1$ and $R_2$. Therefore, we find that the Uhlmann connection is approximately given by
\be
A^U_x(k_x)\approx\frac{i}2f_T\Big(\hat{R}_1 \p_x\hat{R}_2-\hat{R}_2 \p_x\hat{R}_1\Big)\Gamma_{12}
\ee
which is proportional to a fixed matrix $\Gamma_{12}$. Thus $A^U_x(k_x)$ with different $k_x$ are all commuting with each other, which makes the path-ordering product trivial. In this case, one can directly integrate the exponent to find the Uhlmann Wilson loops as
\be
&&W^U\approx\exp\Big[-\frac{i}2\int_0^{2\pi}f_T
\big(\hat{R}_1 \p_x\hat{R}_2-\hat{R}_2 \p_x\hat{R}_1\big)dk_x\,\Gamma_{12}\Big]\nonumber\\
&&=\cos(\pi w_T)-i\sin(\pi w_T)\Gamma_{12}
\ee
where we have defined
\be
w_T=\frac1{2\pi}\int_0^{2\pi}f_T\big(\hat{R}_1 \p_x\hat{R}_2-\hat{R}_2 \p_x\hat{R}_1\big)d k_x
\ee
The quantity $w_T$ can be thought of as a modified winding number of a two component vector $(\hat{R}_1,\,\hat{R}_2)$. Due to the factor of $f_T$, $w_T$ is not quantized to integers. Now one can compute the Uhlmann phase as
\be
\Phi^U=\arg\mbox{Tr}(\rho_0 W^U)=\arg[\cos(\pi w_T)]
\ee
Recall that $f_T\to 1$ as $T \to 0$ and $f_T\to 0$ as $T \to \infty$. Now consider the BBH model with $-1<m<1$ at low $T$, we find that $w_T\approx1$ and thus $\Phi^U=\pi$. Note that $f_T$ also depends on $k_y$. When we vary $k_y$, $f_T$ and $w$ may drop below $1/2$, which makes $\cos(\pi w)$ change sign. This causes the $\Phi^U$ curve to jump somewhere in the $k_y$ axis. In the high $T$ limit, $f_T$ is very small, which in turn makes $w_T\approx0$. This implies $\Phi^U=0$ at high $T$. The critical temperature $T_c$ is determined by $w_T=1/2$, where $\cos(\pi w_T)$ changes sign.

The above approximated calculation of $W^U$ is quite rough, so we cannot determine the Uhlmann overlap and $T_c$ precisely in general.
However, we want to show that the Uhlmann Wilson loop $W^U(k_y)$ can be computed analytically for the special case of $m=0$. When $m=0$, we have $R=\sqrt{2}$, which implies that the dispersion of the two energy bands is completely flat. According to Eq.(\ref{eq-AGa}), the Uhlmann connection becomes quite simple in this case, which is given as
\be
&&A^U_x(k_x,k_y)=\frac{i}{4}f_T\Big(\sin k_y\cos k_x\Gamma_{13}+\cos k_y\cos k_x\Gamma_{23}\nonumber\\
&&\qquad +\sin k_y\sin k_x\Gamma_{14}+\cos k_y\sin k_x\Gamma_{24}+\Gamma_{34}\Big)
\ee
It is more convenient to express $A^U_x$ in a block diagonal form as
\be
A^U_x(k_x,k_y)=\left(
        \begin{array}{cc}
          A_1 & 0 \\
          0 & A_2
        \end{array}
      \right)
\ee
where $A_1$ and $A_2$ are 2 by 2 matrices given as
\be
&&A_1=\frac i4 f_T\Big[-\cos(k_x+k_y)\sigma_1+\sin(k_x+k_y)\sigma_2+\sigma_3\Big]\nonumber\\
&&A_2=-\frac i4 f_T\Big[\cos(k_x-k_y)\sigma_1+\sin(k_x-k_y)\sigma_2+\sigma_3\Big]\nonumber
\ee
Accordingly, $W^U(k_y)$ is also block diagonal as follows
\be
W^U(k_y)=\left(
        \begin{array}{cc}
          W_1 & 0 \\
          0 & W_2
        \end{array}
      \right)
\ee
where $W_a=\tilde{\mathcal{P}}\exp(\oint A_a d k_x)$ for $a=1,2$.

\begin{figure}[t]
\centering
\includegraphics[width=0.7\columnwidth]{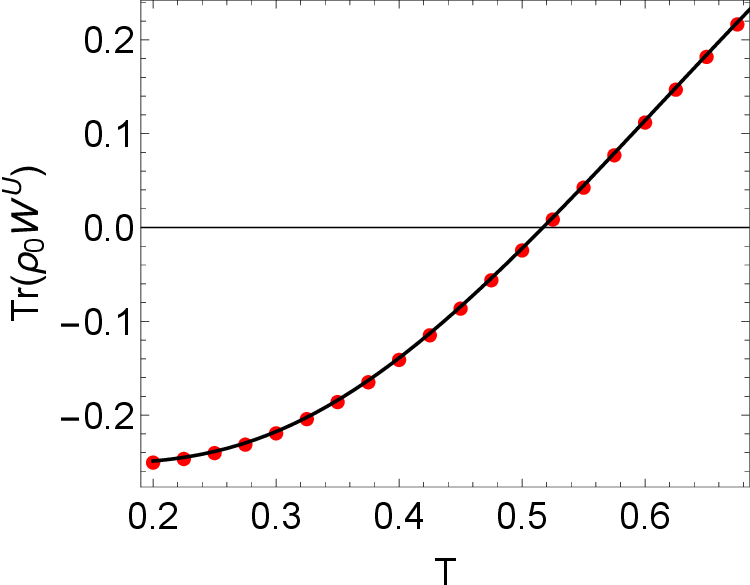}
\caption{The Uhlmann overlap Tr$(\rho_0W^U)$ of the BBH model with $m=0$ as a function of temperature $T$. The black line is the analytical result of Eq.(\ref{eq-overlap}) and the red dot are numerical results.}
\label{fig-cmp}
\end{figure}

Now we focus on the calculation of $W_1$. Making a variable change as $k_x+k_y\to x$, we can then define a Wilson-line operator as
\be
F_1(x)=\tilde{\mathcal{P}}\exp\Big(\int_0^{x} A_1(x') dx'\Big)
\label{eq-F1}
\ee
From the above equation, one can see that $W_1$ is obtained by $W_1=F_1(2\pi)$ and is independent of $k_y$. According to the definition of Eq.(\ref{eq-F1}), the Wilson-line operator $F_1(x)$ satisfies the following
differential equation
\be
\frac{d F_1(x)}{d x}=F_1(x)A_1(x)
\ee
To solve the above equation, we make a gauge transformation as $F_1(x)=F'_1(x) U$ with a unitary matrix $U=e^{-ix\sigma_3/2}$. Then the new Wilson-line operator $F'_1(x)$
satisfies the following equation as
\be
\frac{d F'_1(x)}{d x}=F'_1(x)A'_1(x)
\ee
where $A'_1(x)$ is the gauge transformed Uhlmann connection, which is given by
\be
A'_1(x)&=&UA_1(x)U^{\dag}-\frac{d U}{d x}U^\dag\nonumber\\
&=&\frac i2\Big([\frac{f_T}2+1]\sigma_3-\frac{f_T}2\sigma_1\Big)
\ee
It is easy to see that $A'_1(x)$ is actually independent of $x$. Therefore, it is straightforward to integrate the above equation to find that
\be
W_1&=&F_1(2\pi)=-F'_1(2\pi)\nonumber\\
&=&-\exp\Big[\pi i\Big([\frac{f_T}2+1]\sigma_3-\frac{f_T}2\sigma_1\Big)\Big]
\ee
Similarly, we also find that
\be
W_2=-\exp\Big[\pi i\Big(-[\frac{f_T}2+1]\sigma_3-\frac{f_T}2\sigma_1\Big)\Big]
\ee
With the Uhlmann Wilson-loop in hand, it is easy to compute the Uhlmann overlap as
\be
\mbox{Tr}(\rho_0 W^U)&=&\frac14\Big(\mbox{Tr}W_1+\mbox{Tr}W_2\Big)\nonumber\\
&=&-\cos\Bigg(\pi\sqrt{\frac{f_T^2}{2}+f_T+1}~\Bigg)
\label{eq-overlap}
\ee
To test the above result, we plot Tr$(\rho_0W^U)$ of Eq.(\ref{eq-overlap}) as a function of $T$ together with the numerical results of the Uhlmann overlap (red dots) in Fig. \ref{fig-cmp}. One can see that analytical results exactly agree with the previous numerical results.

For low $T$, one can verify that $3/2<\sqrt{f_T^2/2+f_T+1}<2$, then the above Uhlmann overlap is negative. This gives rise to $\Phi^U=\pi$, which is consistent with the topological phase at $T=0$.
As we increase the temperature, $f_T$ becomes smaller. When the factor $\sqrt{f_T^2/2+f_T+1}$ reaches $3/2$, the Uhlmann overlap will change sign.
Therefore, the critical temperature $T_c$ of the Uhlmann phase is determined by
\be
\sqrt{\frac{f_T^2}{2}+f_T+1}=\frac32
\ee
From the above equation, we find that
\be
T_c=\frac{\sqrt{2}}{\cosh^{-1}(4+\sqrt{14})}\approx 0.517,
\ee
which agrees with $T_c$ curve in the phase diagram of Fig. \ref{fig-Tc}. The BBH model provides one of the rare examples in which the critical temperature of topological transition can be exactly determined.

\section{Conclusion}
\label{sec-con}

In conclusion, we have employed the Uhlmann phase as a finite-temperature topological index for the BBH model. By parallel transporting a mixed state along a closed path using the Uhlmann connection, the resulting state differs from the initial one, and the Uhlmann phase is defined as the phase angle of their overlap. We find that the Uhlmann phase of the BBH model is always quantized to either 0 or $\pi$, allowing it to serve as a topology indicator at finite temperature. Topological phases are marked by abrupt jumps of the Uhlmann phase between these two values, reflecting a modified winding number in the Brillouin zone. These jumps vanish at high temperature, signaling a topological phase transition at a critical temperature $T_c$. We have provided a heuristic explanation for this transition and, in a special case, obtained an analytic expression for $T_c$.

\begin{acknowledgments}
This work is supported by the National Natural Science Foundation of China under Grant No. 11874272.
\end{acknowledgments}

%\bibliography{Ref}

\begin{thebibliography}{33}%
\makeatletter
\providecommand \@ifxundefined [1]{%
 \@ifx{#1\undefined}
}%
\providecommand \@ifnum [1]{%
 \ifnum #1\expandafter \@firstoftwo
 \else \expandafter \@secondoftwo
 \fi
}%
\providecommand \@ifx [1]{%
 \ifx #1\expandafter \@firstoftwo
 \else \expandafter \@secondoftwo
 \fi
}%
\providecommand \natexlab [1]{#1}%
\providecommand \enquote  [1]{``#1''}%
\providecommand \bibnamefont  [1]{#1}%
\providecommand \bibfnamefont [1]{#1}%
\providecommand \citenamefont [1]{#1}%
\providecommand \href@noop [0]{\@secondoftwo}%
\providecommand \href [0]{\begingroup \@sanitize@url \@href}%
\providecommand \@href[1]{\@@startlink{#1}\@@href}%
\providecommand \@@href[1]{\endgroup#1\@@endlink}%
\providecommand \@sanitize@url [0]{\catcode `\\12\catcode `\$12\catcode
  `\&12\catcode `\#12\catcode `\^12\catcode `\_12\catcode `\%12\relax}%
\providecommand \@@startlink[1]{}%
\providecommand \@@endlink[0]{}%
\providecommand \url  [0]{\begingroup\@sanitize@url \@url }%
\providecommand \@url [1]{\endgroup\@href {#1}{\urlprefix }}%
\providecommand \urlprefix  [0]{URL }%
\providecommand \Eprint [0]{\href }%
\providecommand \doibase [0]{http://dx.doi.org/}%
\providecommand \selectlanguage [0]{\@gobble}%
\providecommand \bibinfo  [0]{\@secondoftwo}%
\providecommand \bibfield  [0]{\@secondoftwo}%
\providecommand \translation [1]{[#1]}%
\providecommand \BibitemOpen [0]{}%
\providecommand \bibitemStop [0]{}%
\providecommand \bibitemNoStop [0]{.\EOS\space}%
\providecommand \EOS [0]{\spacefactor3000\relax}%
\providecommand \BibitemShut  [1]{\csname bibitem#1\endcsname}%
\let\auto@bib@innerbib\@empty
%</preamble>
\bibitem [{\citenamefont {Hasan}\ and\ \citenamefont
  {Kane}(2010)}]{Kane_TIRev}%
  \BibitemOpen
  \bibfield  {author} {\bibinfo {author} {\bibfnamefont {M.~Z.}\ \bibnamefont
  {Hasan}}\ and\ \bibinfo {author} {\bibfnamefont {C.L.}\ \bibnamefont
  {Kane}},\ }\bibfield  {title} {\enquote {\bibinfo {title} {Colloquium:
  Topological insulators},}\ }\href@noop {} {\bibfield  {journal} {\bibinfo
  {journal} {Rev. Mod. Phys.}\ }\textbf {\bibinfo {volume} {82}},\ \bibinfo
  {pages} {3045} (\bibinfo {year} {2010})}\BibitemShut {NoStop}%
\bibitem [{\citenamefont {Qi}\ and\ \citenamefont {Zhang}(2011)}]{Zhang_TIRev}%
  \BibitemOpen
  \bibfield  {author} {\bibinfo {author} {\bibfnamefont {Xiao-Liang}\
  \bibnamefont {Qi}}\ and\ \bibinfo {author} {\bibfnamefont {Shou-Cheng}\
  \bibnamefont {Zhang}},\ }\bibfield  {title} {\enquote {\bibinfo {title}
  {Topological insulators and superconductors},}\ }\href@noop {} {\bibfield
  {journal} {\bibinfo  {journal} {Rev. Mod. Phys.}\ }\textbf {\bibinfo {volume}
  {83}},\ \bibinfo {pages} {1057} (\bibinfo {year} {2011})}\BibitemShut
  {NoStop}%
\bibitem [{\citenamefont {Bernevig}\ and\ \citenamefont
  {Hughes}(2013)}]{Bernevig_book}%
  \BibitemOpen
  \bibfield  {author} {\bibinfo {author} {\bibfnamefont {B~A}\ \bibnamefont
  {Bernevig}}\ and\ \bibinfo {author} {\bibfnamefont {T~L}\ \bibnamefont
  {Hughes}},\ }\href@noop {} {\emph {\bibinfo {title} {Topological insulators
  and topological superconductors}}}\ (\bibinfo  {publisher} {Princeton
  University press},\ \bibinfo {address} {Princeton, NJ},\ \bibinfo {year}
  {2013})\BibitemShut {NoStop}%
\bibitem [{\citenamefont {Shen}(2012)}]{ShenTI}%
  \BibitemOpen
  \bibfield  {author} {\bibinfo {author} {\bibfnamefont {Shun-Qing}\
  \bibnamefont {Shen}},\ }\href@noop {} {\emph {\bibinfo {title} {Topological
  Insulators: Dirac Equation in Condensed Matters}}},\ Vol.\ \bibinfo {volume}
  {174}\ (\bibinfo  {publisher} {Springer},\ \bibinfo {year}
  {2012})\BibitemShut {NoStop}%
\bibitem [{\citenamefont {Chiu}\ \emph {et~al.}(2016)\citenamefont {Chiu},
  \citenamefont {Teo}, \citenamefont {Schnyder},\ and\ \citenamefont
  {Ryu}}]{Chiu2016}%
  \BibitemOpen
  \bibfield  {author} {\bibinfo {author} {\bibfnamefont {Ching-Kai}\
  \bibnamefont {Chiu}}, \bibinfo {author} {\bibfnamefont {Jeffrey C.~Y.}\
  \bibnamefont {Teo}}, \bibinfo {author} {\bibfnamefont {Andreas~P}\
  \bibnamefont {Schnyder}}, \ and\ \bibinfo {author} {\bibfnamefont {Shinsei}\
  \bibnamefont {Ryu}},\ }\bibfield  {title} {\enquote {\bibinfo {title}
  {{Classification of topological quantum matter with symmetries}},}\ }\href
  {\doibase 10.1103/RevModPhys.88.035005} {\bibfield  {journal} {\bibinfo
  {journal} {Rev. Mod. Phys.}\ }\textbf {\bibinfo {volume} {88}},\ \bibinfo
  {pages} {035005} (\bibinfo {year} {2016})}\BibitemShut {NoStop}%
\bibitem [{\citenamefont {Thouless}\ \emph {et~al.}(1982)\citenamefont
  {Thouless}, \citenamefont {Kohmoto}, \citenamefont {Nightingale},\ and\
  \citenamefont {den Nijs}}]{TKNN}%
  \BibitemOpen
  \bibfield  {author} {\bibinfo {author} {\bibfnamefont {D.~J.}\ \bibnamefont
  {Thouless}}, \bibinfo {author} {\bibfnamefont {M.}~\bibnamefont {Kohmoto}},
  \bibinfo {author} {\bibfnamefont {M.~P.}\ \bibnamefont {Nightingale}}, \ and\
  \bibinfo {author} {\bibfnamefont {M.}~\bibnamefont {den Nijs}},\ }\bibfield
  {title} {\enquote {\bibinfo {title} {Quantized hall conductance in a
  two-dimensional periodic potential},}\ }\href {\doibase
  10.1103/PhysRevLett.49.405} {\bibfield  {journal} {\bibinfo  {journal}
  {Physical Review Letters}\ }\textbf {\bibinfo {volume} {49}},\ \bibinfo
  {pages} {405--408} (\bibinfo {year} {1982})}\BibitemShut {NoStop}%
\bibitem [{\citenamefont {Hatsugai}(1993)}]{hatsugai1993chern}%
  \BibitemOpen
  \bibfield  {author} {\bibinfo {author} {\bibfnamefont {Yasuhiro}\
  \bibnamefont {Hatsugai}},\ }\bibfield  {title} {\enquote {\bibinfo {title}
  {Chern number and edge states in the integer quantum hall effect},}\
  }\href@noop {} {\bibfield  {journal} {\bibinfo  {journal} {Physical Review
  Letters}\ }\textbf {\bibinfo {volume} {71}},\ \bibinfo {pages} {3697}
  (\bibinfo {year} {1993})}\BibitemShut {NoStop}%
\bibitem [{\citenamefont {Sjoqvist}\ \emph {et~al.}(2000)\citenamefont
  {Sjoqvist}, \citenamefont {Pati}, \citenamefont {Ekert}, \citenamefont
  {Anandan}, \citenamefont {Ericsson}, \citenamefont {Oi},\ and\ \citenamefont
  {Vedral}}]{Sjoqvist00}%
  \BibitemOpen
  \bibfield  {author} {\bibinfo {author} {\bibfnamefont {E}~\bibnamefont
  {Sjoqvist}}, \bibinfo {author} {\bibfnamefont {A~K}\ \bibnamefont {Pati}},
  \bibinfo {author} {\bibfnamefont {A}~\bibnamefont {Ekert}}, \bibinfo {author}
  {\bibfnamefont {J~S}\ \bibnamefont {Anandan}}, \bibinfo {author}
  {\bibfnamefont {M}~\bibnamefont {Ericsson}}, \bibinfo {author} {\bibfnamefont
  {D~K~L}\ \bibnamefont {Oi}}, \ and\ \bibinfo {author} {\bibfnamefont
  {V}~\bibnamefont {Vedral}},\ }\bibfield  {title} {\enquote {\bibinfo {title}
  {Geometric phases for mixed states in interferometry},}\ }\href@noop {}
  {\bibfield  {journal} {\bibinfo  {journal} {Phys. Rev. Lett.}\ }\textbf
  {\bibinfo {volume} {85}},\ \bibinfo {pages} {2845} (\bibinfo {year}
  {2000})}\BibitemShut {NoStop}%
\bibitem [{\citenamefont {Huang}\ and\ \citenamefont {Arovas}(2014)}]{Huang14}%
  \BibitemOpen
  \bibfield  {author} {\bibinfo {author} {\bibfnamefont {Z}~\bibnamefont
  {Huang}}\ and\ \bibinfo {author} {\bibfnamefont {D~P}\ \bibnamefont
  {Arovas}},\ }\bibfield  {title} {\enquote {\bibinfo {title} {Topological
  indices for open and thermal systems via uhlmann phase},}\ }\href@noop {}
  {\bibfield  {journal} {\bibinfo  {journal} {Phys. Rev. Lett.}\ }\textbf
  {\bibinfo {volume} {113}},\ \bibinfo {pages} {076407} (\bibinfo {year}
  {2014})}\BibitemShut {NoStop}%
\bibitem [{\citenamefont {Viyuela}\ \emph
  {et~al.}(2014{\natexlab{a}})\citenamefont {Viyuela}, \citenamefont {Rivas},\
  and\ \citenamefont {Martin-Delgado}}]{Viyuela-1d}%
  \BibitemOpen
  \bibfield  {author} {\bibinfo {author} {\bibfnamefont {O.}~\bibnamefont
  {Viyuela}}, \bibinfo {author} {\bibfnamefont {A.}~\bibnamefont {Rivas}}, \
  and\ \bibinfo {author} {\bibfnamefont {M.~A.}\ \bibnamefont
  {Martin-Delgado}},\ }\bibfield  {title} {\enquote {\bibinfo {title} {Uhlmann
  phase as a topological measure for one-dimensional fermion systems},}\
  }\href@noop {} {\bibfield  {journal} {\bibinfo  {journal} {Phys. Rev. Lett.}\
  }\textbf {\bibinfo {volume} {112}},\ \bibinfo {pages} {130401} (\bibinfo
  {year} {2014}{\natexlab{a}})}\BibitemShut {NoStop}%
\bibitem [{\citenamefont {Viyuela}\ \emph {et~al.}(2015)\citenamefont
  {Viyuela}, \citenamefont {Rivas},\ and\ \citenamefont
  {Martin-Delgado}}]{Viyuela-review}%
  \BibitemOpen
  \bibfield  {author} {\bibinfo {author} {\bibfnamefont {O.}~\bibnamefont
  {Viyuela}}, \bibinfo {author} {\bibfnamefont {A.}~\bibnamefont {Rivas}}, \
  and\ \bibinfo {author} {\bibfnamefont {M.~A.}\ \bibnamefont
  {Martin-Delgado}},\ }\bibfield  {title} {\enquote {\bibinfo {title}
  {Symmetry-protected topological phases at finite temperature},}\ }\href@noop
  {} {\bibfield  {journal} {\bibinfo  {journal} {2D Mater.}\ }\textbf {\bibinfo
  {volume} {2}},\ \bibinfo {pages} {034006} (\bibinfo {year}
  {2015})}\BibitemShut {NoStop}%
\bibitem [{\citenamefont {Budich}\ and\ \citenamefont {Diehl}(2015)}]{Diehl}%
  \BibitemOpen
  \bibfield  {author} {\bibinfo {author} {\bibfnamefont {J~C}\ \bibnamefont
  {Budich}}\ and\ \bibinfo {author} {\bibfnamefont {S}~\bibnamefont {Diehl}},\
  }\bibfield  {title} {\enquote {\bibinfo {title} {Topology of density
  matrices},}\ }\href@noop {} {\bibfield  {journal} {\bibinfo  {journal} {Phys.
  Rev. B}\ }\textbf {\bibinfo {volume} {91}},\ \bibinfo {pages} {165140}
  (\bibinfo {year} {2015})}\BibitemShut {NoStop}%
\bibitem [{\citenamefont {Unanyan}\ \emph {et~al.}(2020)\citenamefont
  {Unanyan}, \citenamefont {Kiefer-Emmanouilidis},\ and\ \citenamefont
  {Fleischhauer}}]{Unanyan20}%
  \BibitemOpen
  \bibfield  {author} {\bibinfo {author} {\bibfnamefont {R}~\bibnamefont
  {Unanyan}}, \bibinfo {author} {\bibfnamefont {M}~\bibnamefont
  {Kiefer-Emmanouilidis}}, \ and\ \bibinfo {author} {\bibfnamefont
  {M}~\bibnamefont {Fleischhauer}},\ }\bibfield  {title} {\enquote {\bibinfo
  {title} {Finite-temperature topological invariant for interacting systems},}\
  }\href@noop {} {\bibfield  {journal} {\bibinfo  {journal} {Phys. Rev. Lett.}\
  }\textbf {\bibinfo {volume} {125}},\ \bibinfo {pages} {215701} (\bibinfo
  {year} {2020})}\BibitemShut {NoStop}%
\bibitem [{\citenamefont {Berry}(1984)}]{Berry}%
  \BibitemOpen
  \bibfield  {author} {\bibinfo {author} {\bibfnamefont {M~V}\ \bibnamefont
  {Berry}},\ }\bibfield  {title} {\enquote {\bibinfo {title} {Quantal phase
  factors accompanying adiabatic changes},}\ }\href@noop {} {\bibfield
  {journal} {\bibinfo  {journal} {Proc. R. Soc. London, Ser. A}\ }\textbf
  {\bibinfo {volume} {392}},\ \bibinfo {pages} {45} (\bibinfo {year}
  {1984})}\BibitemShut {NoStop}%
\bibitem [{\citenamefont {Bardyn}\ \emph {et~al.}(2018)\citenamefont {Bardyn},
  \citenamefont {Wawer}, \citenamefont {Altland}, \citenamefont
  {Fleischhauer},\ and\ \citenamefont {Diehl}}]{Diehl18}%
  \BibitemOpen
  \bibfield  {author} {\bibinfo {author} {\bibfnamefont {C~E}\ \bibnamefont
  {Bardyn}}, \bibinfo {author} {\bibfnamefont {L}~\bibnamefont {Wawer}},
  \bibinfo {author} {\bibfnamefont {A}~\bibnamefont {Altland}}, \bibinfo
  {author} {\bibfnamefont {M}~\bibnamefont {Fleischhauer}}, \ and\ \bibinfo
  {author} {\bibfnamefont {S}~\bibnamefont {Diehl}},\ }\bibfield  {title}
  {\enquote {\bibinfo {title} {Probing the topology of density matrices},}\
  }\href@noop {} {\bibfield  {journal} {\bibinfo  {journal} {Phys. Rev. X}\
  }\textbf {\bibinfo {volume} {8}},\ \bibinfo {pages} {011035} (\bibinfo {year}
  {2018})}\BibitemShut {NoStop}%
\bibitem [{\citenamefont {Uhlmann}(1986)}]{Uhlmann}%
  \BibitemOpen
  \bibfield  {author} {\bibinfo {author} {\bibfnamefont {A}~\bibnamefont
  {Uhlmann}},\ }\bibfield  {title} {\enquote {\bibinfo {title} {Parallel
  transport and ``quantum holonomy'' along density operators},}\ }\href@noop {}
  {\bibfield  {journal} {\bibinfo  {journal} {Rep. Math. Phys.}\ }\textbf
  {\bibinfo {volume} {24}},\ \bibinfo {pages} {229} (\bibinfo {year}
  {1986})}\BibitemShut {NoStop}%
\bibitem [{\citenamefont {Uhlmann}(1991)}]{Uhlmann1}%
  \BibitemOpen
  \bibfield  {author} {\bibinfo {author} {\bibfnamefont {A}~\bibnamefont
  {Uhlmann}},\ }\bibfield  {title} {\enquote {\bibinfo {title} {A gauge field
  governing parallel transport along mixed states},}\ }\href@noop {} {\bibfield
   {journal} {\bibinfo  {journal} {Lett. Math. Phys.}\ }\textbf {\bibinfo
  {volume} {21}},\ \bibinfo {pages} {229} (\bibinfo {year} {1991})}\BibitemShut
  {NoStop}%
\bibitem [{\citenamefont {Viyuela}\ \emph
  {et~al.}(2014{\natexlab{b}})\citenamefont {Viyuela}, \citenamefont {Rivas},\
  and\ \citenamefont {Martin-Delgado}}]{Viyuela-2d}%
  \BibitemOpen
  \bibfield  {author} {\bibinfo {author} {\bibfnamefont {O.}~\bibnamefont
  {Viyuela}}, \bibinfo {author} {\bibfnamefont {A.}~\bibnamefont {Rivas}}, \
  and\ \bibinfo {author} {\bibfnamefont {M.~A.}\ \bibnamefont
  {Martin-Delgado}},\ }\bibfield  {title} {\enquote {\bibinfo {title}
  {Two-dimensional density-matrix topological fermionic phases: Topological
  uhlmann numbers},}\ }\href@noop {} {\bibfield  {journal} {\bibinfo  {journal}
  {Phys. Rev. Lett.}\ }\textbf {\bibinfo {volume} {113}},\ \bibinfo {pages}
  {076408} (\bibinfo {year} {2014}{\natexlab{b}})}\BibitemShut {NoStop}%
\bibitem [{\citenamefont {Mera}\ \emph {et~al.}(2017)\citenamefont {Mera},
  \citenamefont {Vlachou}, \citenamefont {Paunkovic},\ and\ \citenamefont
  {Vieira}}]{Mera17}%
  \BibitemOpen
  \bibfield  {author} {\bibinfo {author} {\bibfnamefont {B}~\bibnamefont
  {Mera}}, \bibinfo {author} {\bibfnamefont {C}~\bibnamefont {Vlachou}},
  \bibinfo {author} {\bibfnamefont {N}~\bibnamefont {Paunkovic}}, \ and\
  \bibinfo {author} {\bibfnamefont {V~R}\ \bibnamefont {Vieira}},\ }\bibfield
  {title} {\enquote {\bibinfo {title} {Uhlmann connection in fermionic systems
  undergoing phase transitions},}\ }\href@noop {} {\bibfield  {journal}
  {\bibinfo  {journal} {Phys. Rev. Lett.}\ }\textbf {\bibinfo {volume} {119}},\
  \bibinfo {pages} {015702} (\bibinfo {year} {2017})}\BibitemShut {NoStop}%
\bibitem [{\citenamefont {He}\ \emph {et~al.}(2018)\citenamefont {He},
  \citenamefont {Guo},\ and\ \citenamefont {Chien}}]{HeChern18}%
  \BibitemOpen
  \bibfield  {author} {\bibinfo {author} {\bibfnamefont {Y}~\bibnamefont {He}},
  \bibinfo {author} {\bibfnamefont {H}~\bibnamefont {Guo}}, \ and\ \bibinfo
  {author} {\bibfnamefont {C~C}\ \bibnamefont {Chien}},\ }\bibfield  {title}
  {\enquote {\bibinfo {title} {Thermal uhlmann-chern number from the uhlmann
  connection for extracting topological properties of mixed states},}\
  }\href@noop {} {\bibfield  {journal} {\bibinfo  {journal} {Phys. Rev. B}\
  }\textbf {\bibinfo {volume} {97}},\ \bibinfo {pages} {235141} (\bibinfo
  {year} {2018})}\BibitemShut {NoStop}%
\bibitem [{\citenamefont {Zhang}\ \emph {et~al.}(2021)\citenamefont {Zhang},
  \citenamefont {Pi}, \citenamefont {He},\ and\ \citenamefont
  {Chien}}]{YH-BHZ}%
  \BibitemOpen
  \bibfield  {author} {\bibinfo {author} {\bibfnamefont {Ye}~\bibnamefont
  {Zhang}}, \bibinfo {author} {\bibfnamefont {Aixin}\ \bibnamefont {Pi}},
  \bibinfo {author} {\bibfnamefont {Yan}\ \bibnamefont {He}}, \ and\ \bibinfo
  {author} {\bibfnamefont {Chih-Chun}\ \bibnamefont {Chien}},\ }\bibfield
  {title} {\enquote {\bibinfo {title} {Comparison of finite-temperature
  topological indicators based on uhlmann connection},}\ }\href {\doibase
  10.1103/PhysRevB.104.165417} {\bibfield  {journal} {\bibinfo  {journal}
  {Phys. Rev. B}\ }\textbf {\bibinfo {volume} {104}},\ \bibinfo {pages}
  {165417} (\bibinfo {year} {2021})}\BibitemShut {NoStop}%
\bibitem [{\citenamefont {Morachis~Galindo}\ \emph {et~al.}(2021)\citenamefont
  {Morachis~Galindo}, \citenamefont {Rojas},\ and\ \citenamefont
  {Maytorena}}]{Galindo21}%
  \BibitemOpen
  \bibfield  {author} {\bibinfo {author} {\bibfnamefont {D}~\bibnamefont
  {Morachis~Galindo}}, \bibinfo {author} {\bibfnamefont {F}~\bibnamefont
  {Rojas}}, \ and\ \bibinfo {author} {\bibfnamefont {J~A}\ \bibnamefont
  {Maytorena}},\ }\bibfield  {title} {\enquote {\bibinfo {title} {Topological
  uhlmann phase transitions for a spin-j particle in a magnetic field},}\
  }\href@noop {} {\bibfield  {journal} {\bibinfo  {journal} {Phys. Rev. A}\
  }\textbf {\bibinfo {volume} {103}},\ \bibinfo {pages} {042221} (\bibinfo
  {year} {2021})}\BibitemShut {NoStop}%
\bibitem [{\citenamefont {Hou}\ \emph {et~al.}(2021)\citenamefont {Hou},
  \citenamefont {Guo},\ and\ \citenamefont {Chien}}]{HouPRA21}%
  \BibitemOpen
  \bibfield  {author} {\bibinfo {author} {\bibfnamefont {X~Y}\ \bibnamefont
  {Hou}}, \bibinfo {author} {\bibfnamefont {H}~\bibnamefont {Guo}}, \ and\
  \bibinfo {author} {\bibfnamefont {C~C}\ \bibnamefont {Chien}},\ }\bibfield
  {title} {\enquote {\bibinfo {title} {Finite-temperature topological phase
  transitions of spin-$j$ systems in uhlmann processes: General formalism and
  experimental protocols},}\ }\href@noop {} {\bibfield  {journal} {\bibinfo
  {journal} {Phys. Rev. A}\ }\textbf {\bibinfo {volume} {104}},\ \bibinfo
  {pages} {023303} (\bibinfo {year} {2021})}\BibitemShut {NoStop}%
\bibitem [{\citenamefont {Hou}\ \emph {et~al.}(2020)\citenamefont {Hou},
  \citenamefont {Gao}, \citenamefont {Guo}, \citenamefont {He}, \citenamefont
  {Liu},\ and\ \citenamefont {Chien}}]{HouPRB20}%
  \BibitemOpen
  \bibfield  {author} {\bibinfo {author} {\bibfnamefont {X~Y}\ \bibnamefont
  {Hou}}, \bibinfo {author} {\bibfnamefont {Q~C}\ \bibnamefont {Gao}}, \bibinfo
  {author} {\bibfnamefont {H}~\bibnamefont {Guo}}, \bibinfo {author}
  {\bibfnamefont {Y}~\bibnamefont {He}}, \bibinfo {author} {\bibfnamefont
  {T}~\bibnamefont {Liu}}, \ and\ \bibinfo {author} {\bibfnamefont {C~C}\
  \bibnamefont {Chien}},\ }\bibfield  {title} {\enquote {\bibinfo {title}
  {Ubiquity of zeros of the loschmidt amplitude for mixed states in different
  physical processes and its implication},}\ }\href@noop {} {\bibfield
  {journal} {\bibinfo  {journal} {Phys. Rev. B}\ }\textbf {\bibinfo {volume}
  {102}},\ \bibinfo {pages} {104305} (\bibinfo {year} {2020})}\BibitemShut
  {NoStop}%
\bibitem [{\citenamefont {Gao}\ and\ \citenamefont {He}(2023)}]{YH-quench}%
  \BibitemOpen
  \bibfield  {author} {\bibinfo {author} {\bibfnamefont {Zhan}\ \bibnamefont
  {Gao}}\ and\ \bibinfo {author} {\bibfnamefont {Yan}\ \bibnamefont {He}},\
  }\bibfield  {title} {\enquote {\bibinfo {title} {Quantum quench dynamics of
  berry and uhlmann phases in topological systems},}\ }\href {\doibase
  10.1103/PhysRevB.108.085126} {\bibfield  {journal} {\bibinfo  {journal}
  {Phys. Rev. B}\ }\textbf {\bibinfo {volume} {108}},\ \bibinfo {pages}
  {085126} (\bibinfo {year} {2023})}\BibitemShut {NoStop}%
\bibitem [{\citenamefont {Benalcazar}\ \emph {et~al.}(2017)\citenamefont
  {Benalcazar}, \citenamefont {Bernevig},\ and\ \citenamefont
  {Hughes}}]{benalcazar2017quantized}%
  \BibitemOpen
  \bibfield  {author} {\bibinfo {author} {\bibfnamefont {Wladimir~A}\
  \bibnamefont {Benalcazar}}, \bibinfo {author} {\bibfnamefont {B~Andrei}\
  \bibnamefont {Bernevig}}, \ and\ \bibinfo {author} {\bibfnamefont {Taylor~L}\
  \bibnamefont {Hughes}},\ }\bibfield  {title} {\enquote {\bibinfo {title}
  {Quantized electric multipole insulators},}\ }\href@noop {} {\bibfield
  {journal} {\bibinfo  {journal} {Science}\ }\textbf {\bibinfo {volume}
  {357}},\ \bibinfo {pages} {61--66} (\bibinfo {year} {2017})}\BibitemShut
  {NoStop}%
\bibitem [{\citenamefont {Okugawa}\ \emph {et~al.}(2019)\citenamefont
  {Okugawa}, \citenamefont {Hayashi},\ and\ \citenamefont
  {Nakanishi}}]{okugawa2019second}%
  \BibitemOpen
  \bibfield  {author} {\bibinfo {author} {\bibfnamefont {Ryo}\ \bibnamefont
  {Okugawa}}, \bibinfo {author} {\bibfnamefont {Shin}\ \bibnamefont {Hayashi}},
  \ and\ \bibinfo {author} {\bibfnamefont {Takeshi}\ \bibnamefont
  {Nakanishi}},\ }\bibfield  {title} {\enquote {\bibinfo {title} {Second-order
  topological phases protected by chiral symmetry},}\ }\href@noop {} {\bibfield
   {journal} {\bibinfo  {journal} {Physical Review B}\ }\textbf {\bibinfo
  {volume} {100}},\ \bibinfo {pages} {235302} (\bibinfo {year}
  {2019})}\BibitemShut {NoStop}%
\bibitem [{\citenamefont {Benalcazar}\ and\ \citenamefont
  {Cerjan}(2022)}]{benalcazar2022chiral}%
  \BibitemOpen
  \bibfield  {author} {\bibinfo {author} {\bibfnamefont {Wladimir~A}\
  \bibnamefont {Benalcazar}}\ and\ \bibinfo {author} {\bibfnamefont
  {Alexander}\ \bibnamefont {Cerjan}},\ }\bibfield  {title} {\enquote {\bibinfo
  {title} {Chiral-symmetric higher-order topological phases of matter},}\
  }\href@noop {} {\bibfield  {journal} {\bibinfo  {journal} {Physical Review
  Letters}\ }\textbf {\bibinfo {volume} {128}},\ \bibinfo {pages} {127601}
  (\bibinfo {year} {2022})}\BibitemShut {NoStop}%
\bibitem [{\citenamefont {Song}\ \emph {et~al.}(2017)\citenamefont {Song},
  \citenamefont {Fang},\ and\ \citenamefont {Fang}}]{Song-2017}%
  \BibitemOpen
  \bibfield  {author} {\bibinfo {author} {\bibfnamefont {Zhida}\ \bibnamefont
  {Song}}, \bibinfo {author} {\bibfnamefont {Zhong}\ \bibnamefont {Fang}}, \
  and\ \bibinfo {author} {\bibfnamefont {Chen}\ \bibnamefont {Fang}},\
  }\bibfield  {title} {\enquote {\bibinfo {title}
  {$(d\ensuremath{-}2)$-dimensional edge states of rotation symmetry protected
  topological states},}\ }\href {\doibase 10.1103/PhysRevLett.119.246402}
  {\bibfield  {journal} {\bibinfo  {journal} {Phys. Rev. Lett.}\ }\textbf
  {\bibinfo {volume} {119}},\ \bibinfo {pages} {246402} (\bibinfo {year}
  {2017})}\BibitemShut {NoStop}%
\bibitem [{\citenamefont {Franca}\ \emph {et~al.}(2018)\citenamefont {Franca},
  \citenamefont {van~den Brink},\ and\ \citenamefont {Fulga}}]{Franca}%
  \BibitemOpen
  \bibfield  {author} {\bibinfo {author} {\bibfnamefont {S.}~\bibnamefont
  {Franca}}, \bibinfo {author} {\bibfnamefont {J.}~\bibnamefont {van~den
  Brink}}, \ and\ \bibinfo {author} {\bibfnamefont {I.~C.}\ \bibnamefont
  {Fulga}},\ }\bibfield  {title} {\enquote {\bibinfo {title} {An anomalous
  higher-order topological insulator},}\ }\href {\doibase
  10.1103/PhysRevB.98.201114} {\bibfield  {journal} {\bibinfo  {journal} {Phys.
  Rev. B}\ }\textbf {\bibinfo {volume} {98}},\ \bibinfo {pages} {201114}
  (\bibinfo {year} {2018})}\BibitemShut {NoStop}%
\bibitem [{\citenamefont {Yang}\ \emph {et~al.}(2024)\citenamefont {Yang},
  \citenamefont {Wang}, \citenamefont {Li},\ and\ \citenamefont
  {Xu}}]{Yang_2024}%
  \BibitemOpen
  \bibfield  {author} {\bibinfo {author} {\bibfnamefont {Yan-Bin}\ \bibnamefont
  {Yang}}, \bibinfo {author} {\bibfnamefont {Jiong-Hao}\ \bibnamefont {Wang}},
  \bibinfo {author} {\bibfnamefont {Kai}\ \bibnamefont {Li}}, \ and\ \bibinfo
  {author} {\bibfnamefont {Yong}\ \bibnamefont {Xu}},\ }\bibfield  {title}
  {\enquote {\bibinfo {title} {Higher-order topological phases in crystalline
  and non-crystalline systems: a review},}\ }\href {\doibase
  10.1088/1361-648X/ad3abd} {\bibfield  {journal} {\bibinfo  {journal} {Journal
  of Physics: Condensed Matter}\ }\textbf {\bibinfo {volume} {36}},\ \bibinfo
  {pages} {283002} (\bibinfo {year} {2024})}\BibitemShut {NoStop}%
\bibitem [{\citenamefont {Kang}\ \emph {et~al.}(2019)\citenamefont {Kang},
  \citenamefont {Shiozaki},\ and\ \citenamefont {Cho}}]{kang2019many}%
  \BibitemOpen
  \bibfield  {author} {\bibinfo {author} {\bibfnamefont {Byungmin}\
  \bibnamefont {Kang}}, \bibinfo {author} {\bibfnamefont {Ken}\ \bibnamefont
  {Shiozaki}}, \ and\ \bibinfo {author} {\bibfnamefont {Gil~Young}\
  \bibnamefont {Cho}},\ }\bibfield  {title} {\enquote {\bibinfo {title}
  {Many-body order parameters for multipoles in solids},}\ }\href@noop {}
  {\bibfield  {journal} {\bibinfo  {journal} {Physical Review B}\ }\textbf
  {\bibinfo {volume} {100}},\ \bibinfo {pages} {245134} (\bibinfo {year}
  {2019})}\BibitemShut {NoStop}%
\bibitem [{\citenamefont {Wheeler}\ \emph {et~al.}(2019)\citenamefont
  {Wheeler}, \citenamefont {Wagner},\ and\ \citenamefont
  {Hughes}}]{wheeler2019many}%
  \BibitemOpen
  \bibfield  {author} {\bibinfo {author} {\bibfnamefont {William~A}\
  \bibnamefont {Wheeler}}, \bibinfo {author} {\bibfnamefont {Lucas~K}\
  \bibnamefont {Wagner}}, \ and\ \bibinfo {author} {\bibfnamefont {Taylor~L}\
  \bibnamefont {Hughes}},\ }\bibfield  {title} {\enquote {\bibinfo {title}
  {Many-body electric multipole operators in extended systems},}\ }\href@noop
  {} {\bibfield  {journal} {\bibinfo  {journal} {Physical Review B}\ }\textbf
  {\bibinfo {volume} {100}},\ \bibinfo {pages} {245135} (\bibinfo {year}
  {2019})}\BibitemShut {NoStop}%
\end{thebibliography}


\end{document}